\documentclass{ws-procs9x6}

\def\d{{\rm d}}
\def\p{I\!\!P}

\def\lr{\left( }
\def\rr{\right) }
\def\le{\left[ }
\def\re{\right] }

\def\beq{\begin{equation}}
\def\eeq{\end{equation}}
\def\bea{\begin{eqnarray}}
\def\eea{\end{eqnarray}}

\begin{document}

\title{From factorization to its breaking in diffractive dijet production}

\author{M.~KLASEN}

\address{Laboratoire de Physique Subatomique et de Cosmologie,
 Universit\'e Joseph Fourier/CNRS-IN2P3, 53 Avenue des Martyrs, 38026
 Grenoble, France \\
E-mail: klasen@lpsc.in2p3.fr}

\maketitle

\abstracts{
When comparing recent experimental data from the H1 and ZEUS Collaborations
at HERA for diffractive dijet production in deep-inelastic scattering (DIS)
and photoproduction with next-to-leading order (NLO) QCD predictions using
diffractive parton densities, good agreement is found for DIS. However, the
dijet photoproduction data are overestimated by the NLO theory, showing that
factorization breaking occurs at this order. While this is expected
theoretically for resolved photoproduction, the fact that the data are
better described by a global suppression of direct {\em and} resolved
contribution by about a factor of two comes as a surprise. We therefore
discuss in some detail the factorization scheme and scale dependence between
direct and resolved contributions and propose a new factorization scheme for
diffractive dijet photoproduction.
}

\section{Diffractive $ep$ Scattering}

\begin{figure}[ht]
 \centerline{\epsfxsize=0.8\textwidth\epsfbox{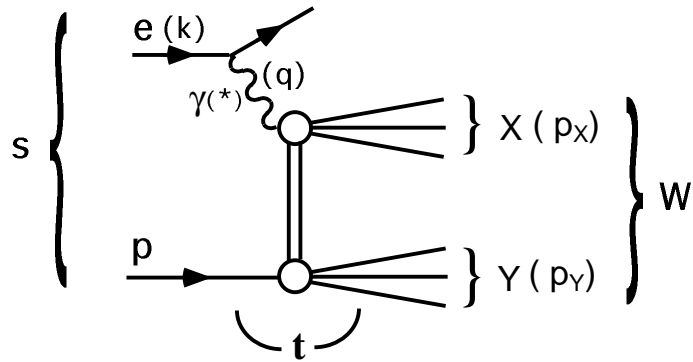}}
 \caption{\label{fig:1}Diffractive scattering process $ep\to eXY$, where
 the hadronic systems $X$ and $Y$ are separated by the largest rapidity
 gap in the final state.}
\end{figure}

It is well known that in high-energy deep-inelastic $ep$-collisions a large
fraction of the observed events are diffractive. These events are defined
experimentally by the presence of a forward-going system $Y$ with
four-momentum $p_Y$, low mass $M_Y$ (in most cases a single proton and/or
low-lying nucleon resonances), small momentum transfer squared
$t=(p-p_Y)^2$, and small longitudinal momentum transfer fraction
$x_{\p}=q(p-p_Y)/qp$ from the incoming proton with four-momentum $p$ to the
system $X$ (see Fig.\ \ref{fig:1}). The presence of a hard scale, as for
%
%
example the photon virtuality $Q^2=-q^2$ in deep-inelastic scattering (DIS)
or the large transverse jet momentum $p_T^{*}$ in the photon-proton
centre-of-momentum frame, should then allow for calculations of the
production cross section for the central system $X$ with the known methods
of perturbative QCD. Under this assumption, the cross section for the
inclusive production of two jets, $e+p \rightarrow e+2~{\rm jets}+X'+Y$,
can be predicted from the well-known formul\ae\ for jet production in
non-diffractive $ep$ collisions, where in the convolution of the partonic
cross section with the parton distribution functions (PDFs) of the proton
the latter ones are replaced by the diffractive PDFs. In the simplest
approximation, they are described by the exchange of a single, factorizable
pomeron/Regge-pole.

\section{Diffractive Parton Distribution Functions}

The diffractive PDFs have been determined by the H1 Collaboration at HERA
from high-precision inclusive measurements of the DIS process $ep\rightarrow
eXY$ using the usual DGLAP evolution equations in leading order (LO) and
next-to-leading order (NLO) and the well-known formula for the inclusive
cross section as a convolution of the inclusive parton-level cross section
with the diffractive PDFs \cite{h1ichep02}. A similar analysis of inclusive
measurements has been published by the ZEUS Collaboration
\cite{Chekanov:2004hy,Abramowicz:2005yc}. A longer discussion of the
extraction of diffractive PDFs can be found elsewhere \cite{Newman:2005wm,%
Martin:2004xw}.

\section{QCD Factorization in Hard Diffraction}

For inclusive diffractive DIS it has been proven by Collins that the formula
referred to above is applicable without additional corrections and that the
inclusive jet production cross section for large $Q^2$ can be calculated in
terms of the same diffractive PDFs \cite{Collins:1997sr}. The proof of this
factorization formula, usually referred to as the validity of QCD
factorization in hard diffraction, may be expected to hold for the direct
part of photoproduction ($Q^2\simeq0$) or low-$Q^2$ electroproduction of
jets \cite{Collins:1997sr}.
However, factorization does not hold for hard processes in diffractive
hadron-hadron scattering. The problem is that soft interactions between the
ingoing two hadrons and their remnants occur in both the initial and final
state. This agrees with experimental measurements at the Tevatron
\cite{Affolder:2000vb}. Predictions of diffractive dijet cross sections for
$p\bar{p}$ collisions as
measured by CDF using the same PDFs as determined by H1 \cite{h1ichep02}
overestimate the measured cross section by up to an order of magnitude
\cite{Affolder:2000vb}. This suppression of the CDF cross section can be
explained by considering the rescattering of the two incoming hadron beams
which, by creating additional hadrons, destroy the rapidity gap
\cite{Kaidalov:2001iz}.

\section{Factorization Breaking in Diffractive Photoproduction}

Processes with real photons ($Q^2 \simeq 0$) or virtual photons with fixed,
but low $Q^2$ involve direct interactions of the photon with quarks from the
proton as well as resolved photon contributions, leading to parton-parton
interactions and an additional remnant jet coming from the photon (for a
review see \cite{Klasen:2002xb}). As already said, factorization should be
valid for direct interactions as in the case of DIS, whereas it is expected
to fail for the resolved process similar as in the hadron-hadron scattering
process. In a two-channel eikonal model similar to the one used to calculate
the suppression factor in hadron-hadron processes \cite{Kaidalov:2001iz},
introducing vector-meson dominated photon fluctuations, a suppression by
about a factor of three for resolved photoproduction at HERA is predicted
\cite{Kaidalov:2003xf}. Such a suppression factor has recently been applied
to diffractive dijet photoproduction \cite{Klasen:2004tz,Klasen:2004qr} and
compared to preliminary data from H1 \cite{h1ichep04} and ZEUS
\cite{zeusichep04}. While at LO no suppression of the resolved contribution
seemed to be necessary, the NLO corrections increase the cross section
significantly, showing that factorization breaking occurs at this order at
least for resolved photoproduction and that a suppression factor $R$ must
be applied to give a reasonable description of the experimental data.

\section{Factorization Scale Dependence for Real Photons}

\begin{figure}[ht]
 \centerline{\epsfxsize=0.8\textwidth\epsfbox{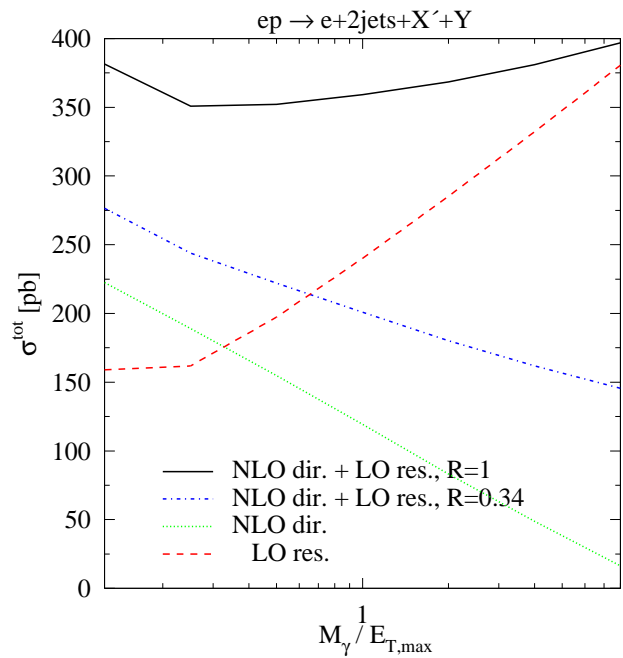}}
 \caption{\label{fig:2}Photon factorization scale dependence of resolved
 (dashed) and direct (dotted) contributions to the diffractive dijet
 photoproduction cross section (full curve). Also shown is the sum of the
 direct and suppressed resolved contribution (dot-dashed curve).}
\end{figure}

As already mentioned elsewhere \cite{Klasen:2004tz,Klasen:2004qr},
describing the factorization breaking in hard photoproduction as well as in
electroproduction at very low $Q^2$ \cite{Klasen:2004ct} by suppressing the
resolved contribution only may be problematic. An indication for this is the
fact that the separation between the direct and the resolved process is
uniquely defined only in LO. In NLO these two processes are related. The
separation depends on the factorization scheme and the factorization scale
$M_{\gamma}$. The sum of both cross sections is the only physically relevant
cross section, which is approximately independent of the factorization
scheme and scale \cite{BKS}. As demonstrated in Fig.\ \ref{fig:2},
multiplying the resolved cross section with the suppression factor $R=0.34$
destroys the correlation of the $M_{\gamma}$-dependence between the direct
and resolved part \cite{Klasen:2004tz,Klasen:2004qr}, and the sum of both
parts has a stronger $M_{\gamma}$-dependence than for the unsuppressed case
($R=1$), where the $M_{\gamma}$-dependence of the NLO direct cross section
is compensated to a high degree by the $M_{\gamma}$-dependence of the
LO resolved part. \\

The introduction of the resolved cross section is dictated by perturbation
theory. At NLO, collinear singularities arise from the photon initial state,
which are absorbed at the factorization scale into the photon PDFs. This way
the photon PDFs become $M_{\gamma}$-dependent. The equivalent
$M_{\gamma}$-dependence, just with the opposite sign, is left in the NLO
corrections to the direct contribution. With this knowledge, it is obvious
that we can obtain a physical cross section at NLO, {\it i.e.} the
superposition of the NLO direct and LO resolved cross section, with a
suppression factor $R<1$ and no $M_{\gamma}$-dependence left, if we also
multiply the $\ln M_{\gamma}$-dependent term of the NLO correction to the
direct contribution with the same suppression factor as the resolved cross
section. We are thus led to the theoretical conclusion that, contrary to
what one may expect, not {\em all} parts of the direct contribution
factorize. Instead, the {\em initial state} singular part appearing beyond
LO breaks factorization even in direct photoproduction, presumably through
soft gluon attachments between the proton and the collinear quark-antiquark
pair emerging from the photon splitting. This would be in agreement with the
non-cancellation of initial state singularities in diffractive hadron-hadron
scattering \cite{Collins:1997sr}.

\section{The Transition Region of Virtual Photoproduction}

We now present the special form of the $\ln M_{\gamma}$-term in
the NLO direct contribution and demonstrate that the $M_{\gamma}$-dependence
of the physical cross section cancels to a large extent in the same way as
in the unsuppressed case ($R=1$). These studies can be done for
photoproduction ($Q^2 \simeq 0$) as well as for electroproduction with
fixed, small $Q^2$. Since in electroproduction the initial-state singularity
in the limit $Q^2 \rightarrow 0$ is more directly apparent than for the
photoproduction case, we shall consider in this contribution the low-$Q^2$
electroproduction case just for demonstration. This diffractive dijet cross
section has been calculated recently \cite{Klasen:2004ct}.

A consistent factorization scheme for low-$Q^2$ virtual photoproduction
has been defined and the full (direct and resolved) NLO corrections for
inclusive dijet production have been calculated in \cite{Klasen:1997jm}. In
this work we adapt this inclusive NLO calculational framework to diffractive
dijet production at low-$Q^2$ in the same way as in \cite{Klasen:2004ct},
except that we multiply the $\ln M_{\gamma}$-dependent terms as well as the
resolved contributions with the same suppression factor $R=0.34$, as an
example, as in our earlier work \cite{Klasen:2004tz,Klasen:2004qr,%
Klasen:2004ct}. The exact value of this suppression factor may change in the
future, when better data for photoproduction and low-$Q^2$ electroproduction
have been analyzed. We present the $\ln M_{\gamma}$-dependence of the partly
suppressed NLO direct and the fully suppressed NLO resolved cross section
$\d\sigma/\d Q^2$ and their sum for the lowest $Q^2$ bin. \\

The NLO corrections for virtual jet photoproduction have been implemented in
the NLO Monte Carlo program JET\-VIP \cite{Potter:1999gg} and adapted to
diffractive dijet
production in \cite{Klasen:2004ct}. The subtraction term, which is absorbed
into the PDFs of the virtual photon $f_{a/\gamma}(x_\gamma,M_{\gamma})$, can
be found in \cite{Klasen:2005dq}. The main term is proportional to
$\ln(M_{\gamma}^2/Q^2)$ times the splitting function
\beq
 P_{q_i \leftarrow \gamma}(z) = 2 N_c Q_i^2 \frac{z^2+(1-z)^2}{2},
 \label{eq:2}
\eeq
where $z=p_1p_2/p_0q \in [x;1]$ and $Q_i$ is the fractional charge of the
quark $q_i$. $p_1$ and $p_2$ are the momenta of the two outgoing jets, and
$p_0$ and $q$ are the momenta of the ingoing parton and virtual photon,
respectively. Since $Q^2=-q^2 \ll M_{\gamma}^2$, the subtraction term is
large and is therefore resummed by the DGLAP evolution equations for the
virtual photon PDFs. After this subtraction, the finite term
$M(Q^2)_{\overline{\rm MS}}$, which remains in the matrix element for the
NLO correction to the direct process \cite{Klasen:1997jm}, has the same 
$M_{\gamma}$-dependence as the subtraction term, {\it i.e.} $\ln M_{\gamma}$
is multiplied with the same factor. As already mentioned, this yields the
$M_{\gamma}$-dependence before the evolution is turned on. In the usual
non-diffractive dijet photoproduction these two $M_{\gamma}$-dependences
cancel, when the NLO correction to the direct part is added to the LO
resolved  cross section \cite{BKS}. Then it is obvious that the approximate
$M_{\gamma}$-independence is destroyed, if the resolved cross section is
multiplied by a suppression factor $R$ to account for the factorization
breaking in the experimental data. To remedy this deficiency, we propose to
multiply the $\ln M_{\gamma}$-dependent term in $M(Q^2)_{\overline{\rm MS}}$
with the same suppression factor as the resolved cross section. This is done
in the following way: we split $M(Q^2)_{\overline{\rm MS}}$ into two terms
using the scale $p_T^{*}$ in such a way that the term containing the slicing
parameter $y_s$, which was used to separate the initial-state singular
contribution, remains unsuppressed. In particular, we replace the finite
term after the subtraction by
\bea
 M(Q^2,R)_{\overline{\rm MS}} &=& \le-\frac{1}{2N_c} P_{q_i\leftarrow
 \gamma}(z)\ln\lr\frac{M_{\gamma}^2 z}{p_T^{*2}(1-z)}\rr+\frac{Q_i^2}{2}\re
 R \nonumber \\
 && \ -\frac{1}{2N_c} P_{q_i\leftarrow\gamma}(z)
 \ln\lr\frac{p_T^{*2}}{zQ^2+y_s s}\rr,\label{eq:4}
\eea
where $R$ is the suppression factor. This expression coincides with the
finite term after subtraction (see Ref.\ \cite{Klasen:2005dq}) for $R=1$, as
it should, and leaves the second term in Eq.\ (\ref{eq:4}) unsuppressed. In
Eq.\ (\ref{eq:4}) we have suppressed in addition to $\ln(M_{\gamma}^2/
p_T^{*2})$ also the $z$-dependent term $\ln (z/(1-z))$, which is specific to
the $\overline{\rm MS}$ subtraction scheme as defined in
\cite{Klasen:1997jm}. The second term in Eq.\ (\ref{eq:4}) must be left in
its original form, {\it i.e.} being unsuppressed, in order to achieve the
cancellation of the slicing parameter ($y_s$) dependence of the complete NLO
correction in the limit of very small $Q^2$ or equivalently very large $s$.
It is clear that the suppression of this part of the NLO correction
to the direct cross section will change the full cross section only very
little as long as we choose $M_{\gamma} \simeq p_T^{*}$. The first term in
Eq.\ (\ref{eq:4}), which has the suppression factor $R$, will be denoted by
${\rm DIR}_{\rm IS}$ in the following.

To study the left-over $M_{\gamma}$-dependence of the physical cross
section, we have calculated the diffractive dijet cross section with the
same kinematic constraints as in the H1 experiment \cite{Schatzel:2004be}. 
Jets are defined by the CDF cone algorithm with jet radius equal to one and
asymmetric cuts for the transverse momenta of the two jets required for
infrared stable comparisons with the NLO calculations \cite{Klasen:1995xe}.
The original H1 analysis actually used a symmetric cut of 4 GeV on the
transverse momenta of both jets \cite{Adloff:2000qi}. The data have,
however, been reanalyzed for asymmetric cuts \cite{Schatzel:2004be}. 

For the NLO resolved virtual photon predictions, we have used the PDFs SaS1D
\cite{Schuler:1996fc} and transformed them from the DIS$_{\gamma}$ to the 
$\overline{\rm MS}$ scheme as in Ref.\ \cite{Klasen:1997jm}. If not stated
otherwise, the renormalization and factorization scales at the pomeron and
the photon vertex are equal and fixed to $p_T^{*} = p_{T,jet1}^{*}$. We
include four flavors, {\it i.e.} $n_f=4$ in the formula for $\alpha_s$ and
in the PDFs of the pomeron and the photon. With these assumptions we have
calculated the same cross section as in our previous work
\cite{Klasen:2004ct}. First we investigated how the cross section
$\d\sigma/\d Q^2$ depends on the factorization scheme of the PDFs for the
virtual photon, {\it i.e.} $\d\sigma/\d Q^2$ is calculated for the choice
SaS1D and SaS1M. Here $\d\sigma/\d Q^2$ is the full cross section (sum of
direct and resolved) integrated over the momentum and rapidity ranges as in
the H1 analysis. The results, shown in Fig.\ 2 of Ref.\ \cite{Klasen:2005dq},
demonstrate that the choice of the factorization scheme
of the virtual photon PDFs has negligible influence on $\d\sigma/\d Q^2$
for all considered $Q^2$. The predictions agree reasonably well with the
preliminary H1 data \cite{Schatzel:2004be}. 

\begin{figure}[ht]
 \centerline{\epsfxsize=0.8\textwidth\epsfbox{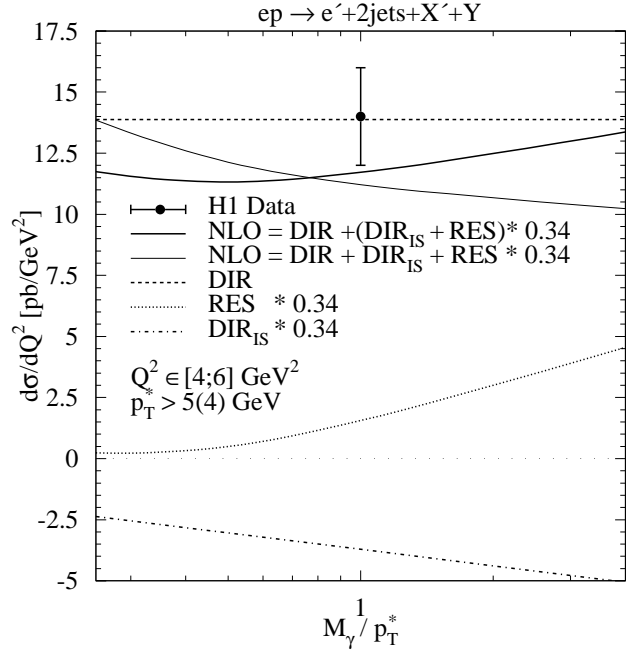}}
 \caption{\label{fig:3}Photon factorization scale dependence of resolved
 and direct contributions to $\d\sigma/\d Q^2$ together with their
 weighted sums for (i) suppression of the resolved cross section and for
 (ii) additional suppression of DIR$_{\rm IS}$, using SaS1D virtual
 photon PDFs.}
\end{figure}

We now turn to the $M_{\gamma}$-dependence of the cross section with a
suppression factor for DIR$_{\rm IS}$. To show this dependence for the two
suppression mechanisms, (i) suppression of the resolved cross section only
and (ii) additional suppression of the DIR$_{\rm IS}$ term as defined in
Eq.\ (\ref{eq:4}) in the NLO correction of the direct cross section, we
consider $\d\sigma/\d Q^2$ for the lowest $Q^2$-bin, $Q^2\in [4,6]$ GeV$^2$.
In Fig.\ \ref{fig:3}, this cross section
is plotted as a function of $\xi=M_{\gamma}/p_T^{*}$ in the range $\xi\in
[0.25;4]$ for the cases (i) (light full curve) and (ii) (full curve). We see
that the cross section for case (i) has an appreciable $\xi$-dependence in
the considered $\xi$ range of the order of $40\%$, which is caused by the
suppression of the resolved contribution only. With the additional
suppression of the DIR$_{\rm IS}$ term in the direct NLO correction, the
$\xi$-dependence of $\d\sigma/\d Q^2$ is reduced to approximately less
than $20\%$, if we compare the maximal and the minimal value of $\d\sigma/
\d Q^2$ in the considered $\xi $ range. The remaining $\xi $-dependence is
caused by the NLO corrections to the suppressed resolved cross section and
the evolution of the virtual photon PDFs. How the compensation of the
$M_{\gamma}$-dependence between the suppressed resolved contribution and the
suppressed direct NLO term works in detail is exhibited by the dotted and
dashed-dotted curves in Fig.\ \ref{fig:3}. The suppressed resolved
term increases and the suppressed direct NLO term decreases by approximately
the same amount with increasing $\xi$. In addition we show also $\d\sigma/
\d Q^2$ in the DIS theory, {\it i.e.} without subtraction of any $\ln Q^2$
terms (dashed line). Of course, this cross section must be independent of
$\xi$. This prediction agrees very well with the experimental point, whereas
the result for the subtracted and suppressed theory (full curve) lies
slightly below. We notice, that for $M_{\gamma}=p^{*}_T$ the additional
suppression of DIR$_{\rm IS}$ has only a small effect. It increases
$\d\sigma/\d Q^2$ by $5\%$ only \cite{Klasen:2005cz,Bruni:2005eb}.

\section{Conclusion}

When comparing experimental data from the H1 and ZEUS Collaborations at HERA
for diffractive dijet production in DIS and photoproduction with NLO QCD
predictions using diffractive parton densities from H1 and ZEUS, good
agreement is found for DIS assuming the H1 diffractive PDFs. However, the
dijet photoproduction data are overestimated by the NLO theory, showing that
factorization breaking occurs at this order. While this is expected
theoretically for resolved photoproduction, the fact that the data are
better described by a global suppression of direct {\em and} resolved
contribution by about a factor of two comes as a surprise. We have
therefore discussed in some detail the factorization scheme and scale
dependence between direct and resolved contributions and proposed a new
factorization scheme for diffractive dijet photoproduction.

\section*{Acknowledgments}
The author thanks the organizers of the Ringberg workshop on {\em New Trends
in HERA Physics 2005} for the kind invitation, G.\ Kramer for his continuing
collaboration, and the {\em Comit\'e de Financement des Projets de Physique
Th\'eorique de l'IN2P3} for financial support.

\end{document}